# AxoNet: an AI-based tool to count retinal ganglion cell axons

Matthew D. Ritch[1], Bailey G. Hannon[2], A. Thomas Read[1], Andrew J. Feola[1,3], Grant A. Cull[4], Juan Reynaud[4], John C. Morrison[5], Claude F. Burgoyne[4], Machelle T. Pardue[1,3], C. Ross Ethier[1,2]

*Abstract—* *Goal*: In this work, we develop a robust, extensible tool to automatically and accurately count retinal ganglion cell axons in images of optic nerve tissue from various animal models of glaucoma. *Methods*: The U-Net convolutional neural network architecture was adapted to learn pixelwise axon count density estimates, which were then integrated over the image area to determine axon counts. The tool, termed AxoNet, was trained and evaluated using a dataset containing images of optic nerve regions randomly selected from complete cross sections of intact rat optic nerves and manually annotated for axon count and location. Both control and damaged optic nerves were used. This rat-trained network was then applied to a separate dataset of non-human primate (NHP) optic nerve images. AxoNet was then compared to two existing automated axon counting tools, AxonMaster and AxonJ, using both datasets. *Results*: AxoNet outperformed the existing tools on both the rat and NHP optic nerve datasets as judged by mean absolute error, $R^2$ values when regressing automated vs. manual counts, and Bland-Altman analysis. *Conclusion*: The proposed tool allows for accurate quantification of axon numbers as a measure of glaucomatous damage. AxoNet is robust to variations in optic nerve tissue damage extent, image quality, and species of mammal. *Significance*: The deep learning method does not rely on hand-crafted image features for axon recognition. Therefore, this approach is not species-specific and can be extended to quantify additional optic nerve features. It will aid evaluation of optic nerve changes in glaucoma and potentially other neurodegenerative diseases.

Paper submitted for review on X/XX/2019. We have greatly benefitted from stimulating discussions with Prof. Ernst Tamm and Dr. Sebastian Koschade. We acknowledge the following funding sources: National Institutes of Health (Bethesda, MD): R01 EY025286 (CRE), 5T32 EY007092-32 (BGH), R01 EY010145 (JCM), P30 EY010572 (JCM), RR&D Service Career Development Award (RX002342; AJF) Research to Prevent Blindness (New York, NY) (JCM), and Georgia Research Alliance (CRE).

Affiliations:
1. Wallace H. Coulter Department of Biomedical Engineering, Georgia Institute of Technology and Emory University, Atlanta, GA, USA.
2. George W. Woodruff School of Mechanical Engineering, Georgia Institute of Technology, Atlanta, GA, USA.
3. Center for Visual and Neurocognitive Rehabilitation, Atlanta VA Healthcare System, Atlanta, GA, USA.
4. Devers Eye Institute, Legacy Research Institute, Portland, OR, USA.
5. Casey Eye Institute, Oregon Health & Science University, Portland, OR, USA.

Prof. C. Ross Ethier is the corresponding author for this work (correspondence e-mail: ross.ethier@bme.gatech.edu).

## I. INTRODUCTION

Glaucoma is the leading cause of irreversible blindness worldwide [1, 2], and thus is a significant research focus. This optic neuropathy is characterized by degeneration and loss of retinal ganglion cells (RGCs), which carry visual signals from the retina to the brain. Therefore, an important outcome measure in studying glaucomatous optic neuropathy, particularly in animal models of the disease, is the number and appearance of RGC axons comprising the optic nerve [3, 4], usually evaluated from images of optic nerve cross sections. Using images obtained by light microscopy is known to result in an axon count underestimation of around 30% relative to counts from images obtained by transmission electron microscopy [5, 6]. However, light microscopy is widely used to count optic nerve axons because of its lower cost and favorable time requirements for tissue preparation. Therefore, in this work we focus on axon counting in optic nerve images generated by light microscopy.

Manual counting is the gold standard approach to quantifying RGC axons, but is extremely labor-intensive, since RGC axon numbers in healthy nerves range from the tens of thousands in mice, to more than a million in humans [7]. Further complicating axon quantification is the fact that axon appearance can be highly variable. For example, in the healthy nerve, most axons are characterized by a clear central axoplasmic core and a darker myelin sheath; following previous work [5, 8], we will refer to such an appearance as "normal"[1]. However, in damaged nerves (and even occasionally in ostensibly heathy nerves), other axon appearances occur, such as an incomplete myelin sheath and/or a darker axoplasmic region. Such variability further increases the time needed for axon counting, since the person doing the counting often needs to decide whether a given feature is (or is not) an axon.

To reduce the time-intensive counting process, various techniques have been developed for assessing axon counts and/or optic nerve damage, including: semi-quantitative, sub-sampling, semi-automated, and automated counting. In the semi-quantitative approach, scores based on a damage grading scale are assigned to optic nerves by different trained observers, and then averaged [8, 9]. While this method is capable of quickly capturing whole-nerve changes, it is subjective and requires scorers who have significant experience and training. Sub-sampling is the process of estimating axon loss by manually counting smaller regions of the nerve using either targeted or random sampling and then extrapolating to the whole nerve or providing an RGC axon count per area measurement [5]. Sub-sampling is faster than full manual counting, but it is still labor-intensive and can be poorly suited to analyzing nerves with regional patterns of axonal loss [9]. Koschade et al. have recently presented an elegant stereological sub-sampling method that eliminates the bias that can occur in sub-sampling, but still requires manual axon counting in 5-10% of the full nerve area [10]. While this is feasible in animals with fewer axons per optic nerve like the mouse, counting this proportion may be prohibitive for animals with more axons per optic nerve, as in primates. Semi-automated axon counting methods use algorithmic axon segmentation techniques involving hyperparameters such as intensity thresholds which are manually tuned for individual sub-images [11]. These methods are faster than manual counting and more thorough than qualitative or sub-sampling methods, but still require extensive human direction and time. Because of these limitations, there has been a push to develop fully automated counting tools.

---

[1] Here and throughout we place the term "normal" in quotes; as will be discussed in more detail below, an "abnormal" appearance does not necessarily imply non-functionality, and it is important to keep this distinction in mind.

Two of the most used automated counting tools are AxonMaster [12] and AxonJ [13]. Both tools are designed to count "normal"-appearing axons, i.e. axons with a clear cytoplasmic core and a dark myelin sheath [5, 8]. They use dynamic thresholding techniques to segment axonal interiors from myelin and other optic nerve features. While these tools are faster and provide more detail than sub-sampling methods, they also suffer limitations. For example, they are not easily extensible to counting features other than "normal"-appearing axons. Further, the two automated counting packages that currently exist were each developed for a specific animal species, and due to inter-species differences, it is not clear how accurate these approaches are for other species. Specifically, AxonMaster [12] and AxonJ [13] were calibrated and validated for use in non-human primate (NHP) and mouse models of glaucoma, respectively. Recently, AxonMaster has been applied to count RGCs in healthy and damaged tree shrew optic nerves [14], but it has yet to be validated in this animal model. Our preliminary testing using these packages suggested that they are also sensitive to image quality, tissue staining intensity, and nerve damage extent in images of rat optic nerves (see below).

Our goal was thus to create axon-counting software to overcome the above limitations, i.e. software which was robust to image quality and staining intensity, which could be used in multiple animal models of glaucoma, and which was extensible to quantification of features other than "normal"-appearing axons. Our approach to building this software, which we refer to as AxoNet, was an adaptation of the U-Net convolutional neural network architecture developed by Ronnenberger et al. [15] applied to the count density learning approach of Lempitsky et al. [16].

We used a dataset of manually annotated rat optic nerve images for developing and training AxoNet (detailed below). The rat is a widely used animal model for glaucoma research and displays retinal structural changes and loss of RGC axons similar to those observed in the human pathology [17]. We then applied our software to the dataset of NHP optic nerve images which was used to validate AxonMaster by Reynaud et al. [12]. Below we present the detailed methodology of the dataset and software construction used to develop AxoNet, as well as a comparison of AxoNet's automated counting results to those of AxonMaster and AxonJ. We have packaged AxoNet into a user friendly open source plugin for the widely-used ImageJ image processing platform [18], as described in greater detail below.

## II. METHODS

### A. Rat Optic Nerve Dataset
*1) Animals*

This study used twenty-seven optic nerves from fourteen (12 male and 2 female) Brown Norway rats aged 3 to 13 months (Charles River Laboratories, Inc., Wilmington, MA). All procedures were approved by the Institutional Animal Care and Use Committee at the Atlanta Veterans Affairs Medical Center and Georgia Institute of Technology. Rats used in this study had various degrees of optic nerve health. Each animal had one eye with experimental glaucoma induced unilaterally by either microbead injection (12 animals) [19-21] or hypertonic saline injection (2 animals) [22]. Optic nerves in the resulting dataset ranged from ostensibly normal to severely damaged due to ocular hypertension.

*2) Tissue Processing and Imaging*

Animals were euthanized via $CO_2$ and the eyes were enucleated. The optic nerves were transected with micro scissors close (<1 mm) to the posterior scleral surface. Optic nerves were then placed in Karnovsky's fixative, post-fixed in osmium tetroxide, dehydrated in an ethanol series, infiltrated and embedded in araldite-epon resin (EMS, Hatfield, PA). Semithin sections of 0.5 μm thickness were cut on a Leica UC7 Ultramicrotome (Leica Microsystems, Buffalo Grove, IL) and stained with 1% toluidine blue. They were imaged with a Leica DM6 B microscope (Leica Microsystems, Buffalo Grove, IL) using a 63x lens and 1.6x multiplier for a total magnification of 100x. A z-stack tile scan of the entire nerve was taken and the optimally focused image within each z-stack tile was selected using the "find best focus" feature in the LAS-X software (Leica Microsystems, Buffalo Grove, IL). Contrast was then adjusted for each tile by maximizing grey-value variance.

*3) Annotated Dataset Construction*

To train the AxoNet algorithm, it was necessary to create a dataset of rat optic nerve images in which axons had been identified. For this purpose, 12 x 12 μm sub-images were randomly selected from the full 27 nerves, producing a dataset of 1474 partial optic nerve images, with a minimum of 20 sub-images selected from each nerve. Image resolution was 17 pixels per μm. Selected sub-images varied in image quality and contrast, and were from optic nerve sections that varied in tissue staining intensity and degree of nerve damage (Figure 1). Four trained counters manually annotated "normal"-appearing axons in 1184 sub-images, where a "normal" axon was defined as a structure with an intact and continuous myelin sheath, a homogenous light interior, and absence of obvious swelling or shrinkage [5, 8]. Each counter annotated one point per axon at the axon's approximate center. The remaining 290 sub-images were annotated by the agreement of two counters. Axons with any abnormal morphology were not annotated. Counters were instructed to count axons which were fully inside the frame of the image or which intersected either the left or top image border and lay more than halfway within the image borders. Manual annotations were made using Fiji's Cell Counter plugin [23], which recorded the spatial location of each axon marked within the image. There was good agreement between manual counts for most sub-images (Figure 2).

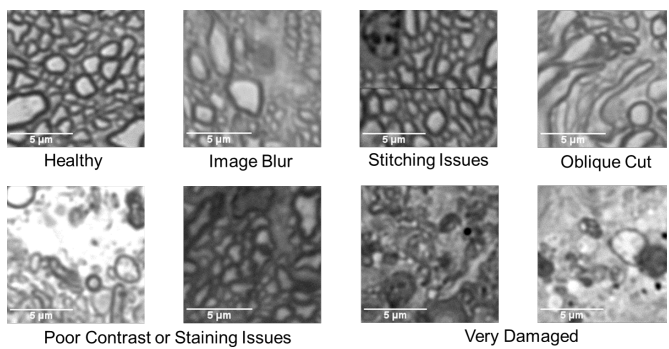

Fig. 1. Rat Dataset Image Variety. A representative set of images from the rat optic nerve image dataset is shown. These images include a range of nerve health, variations in sample processing quality, and in image acquisition contrast and quality.

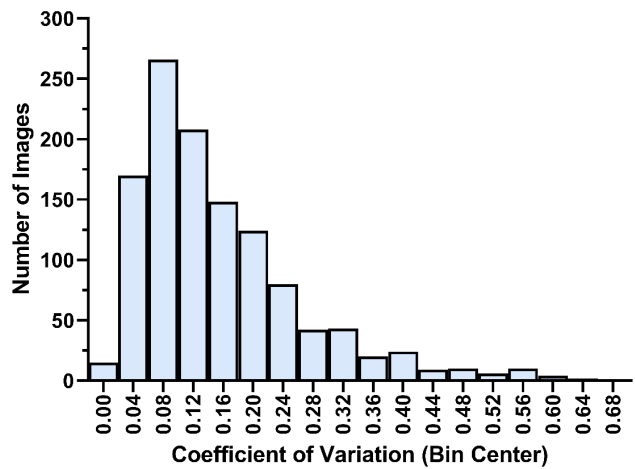

Fig. 2. Histogram of Manual Count Variability for Rat Dataset. Variability between counters is expressed as the coefficient of variation (standard deviation of the manual count divided by the mean of the manual count for each image). The median coefficient of variation was 0.12, indicating good general agreement between manual counters.

These manual annotations were then used to create a "ground truth" axon count density matrix for each sub-image, $D$, in which the $(i,j)^{th}$ entry in the matrix was defined as

$$D(i,j) = \frac{1}{K}\sum_{k=1}^{K} c_k(i,j), \quad (1)$$

where

$$c_k(i,j) = \begin{cases} 1, \text{if the } (i,j)^{th} \text{ pixel was annotated by the k}^{th} \text{ counter} \\ 0, \quad \text{otherwise} \end{cases}$$

and $K$ was the number of counters for the sub-image in question. Note that the dimensions of $D$ equaled the dimensions (in pixels) of the corresponding sub-image. Entries in $D$ were then distributed ("blurred") according to $D_{dist} = \mathcal{G}(D)$, where $\mathcal{G}$ is a Gaussian blur operator with σ=8 and filter size of 33 pixels, chosen empirically to distribute the annotated density values $D(i,j)$ over the full axon. This operation resulted in some of the annotated density values being distributed outside the edges of the original sub-image; we therefore applied a multiplicative correction factor to all entries in $D_{dist}$ so that the integral of the final density values over the entire sub-image equaled the manual count, i.e. so that $\sum D_{dist}(i,j) = \sum D(i,j)$, where the sum was carried out over all entries in the matrix. The resulting ground truth matrix $D_{dist}$ provided the spatial distribution of axon count density over the full sub-image, which when summed over all entries, produces the ground truth axon count for the full sub-image or the average count from all experts for that sub-image.

*4) Dataset Subdivisions*

The dataset was randomly divided into training, validation, and testing image subsets following a 60%-20%-20% split [24]. AxoNet was trained using the training subset. The validation subset was used to optimize AxoNet's architecture and hyperparameters as well as to construct axon count correction equations, as was done using the calibration set in Reynaud et al. [12] and as described below. Finally, the testing subset was used for final evaluation of tool performance.

*B. NHP Dataset*

We then evaluated the performance of AxoNet on optic nerve sub-images from NHPs with experimental glaucoma. This dataset had been previously annotated using a semi-automated manual method and used to develop one of the existing automated axon counting tools, AxonMaster, as described in Reynaud et al. [12].

NHP dataset images were randomly divided into validation and testing subsets following a 50%-50% split to match the even proportion of images in the validation and testing subsets of our rat dataset. The validation subset was used to construct axon count correction equations, as was done using the calibration set in Reynaud et al. [12] and as described below. The testing subset was used for final evaluation of performance for each tool.

## C. AxoNet Development

### 1) Implementation and Network Architecture

We implemented a U-Net based encoder/decoder architecture similar to the original architecture developed by Ronnenberger et al. [15]. Specifically, we reduced the number of filters in our convolutional layers by a factor of two, resulting in a feature depth at each layer half of that in the original architecture. The reduction in filter numbers improved performance in terms of increased count accuracy, reduced the danger of overfitting, and decreased time to achieve model convergence. We used a rectified linear unit (ReLU) instead of a sigmoid activation for the final layer, indicated by the red arrow in Figure 3. The change in the final layer allowed us to regress the ground truth pixelwise count density function instead of predicting cell segmentation. We also included padding on all convolutional layers so that feature arrays would not shrink after each convolution. This network was implemented in Python (Version 3.7.3, Python Software Foundation) using Keras [25] and Tensorflow [26]. All images were normalized to have pixel values in the range of -1.0 to 1.0. The network was trained for 500 epochs at 100 steps per epoch with a batch size of 1 image per step and a learning rate of 10-4. Our modified architecture was developed iteratively by training on the training subset of the rat dataset and evaluating on the validation subset of the rat dataset. Validation performance was used to compare architectures until performance stopped improving.

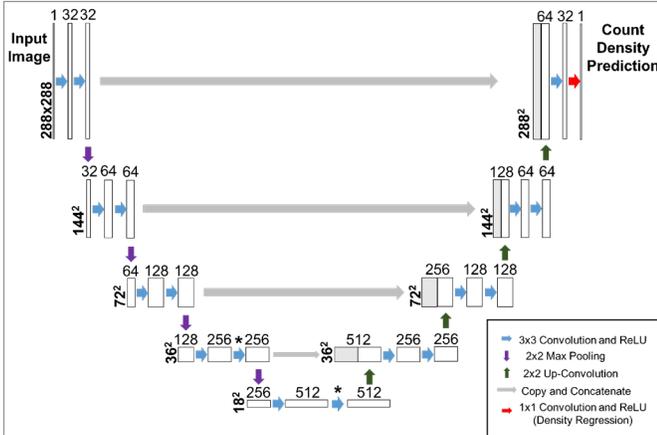

Fig. 3. U-Net Architecture. A visual representation of our adapted U-Net convolutional neural network architecture, with the encoding branch on the left and the decoding branch on the right. Each box represents the output array of one of the network's convolutions, which are represented by colored arrows. The bold numbers to the left of the boxes indicate the row and column sizes of the feature array at those layers. The numbers above the boxes indicate the feature depth of each layer, which is the third dimension of the feature array at that layer. Numbers in the layer operations key indicate the size of that operation's sliding window. Products of feature concatenation are indicated by two boxes sharing a border with the concatenated box in grey. The asterisk indicates dropout with rate = 0.5 applied after convolution. ReLU is an abbreviation for Rectified Linear Unit. Figure adapted from Ronnenberger et al. [15].

### 2) Training

We used the Adam optimizer [27] to minimize a mean squared error loss function evaluated between ground truth and predicted count density function estimates for each image as follows:

$$L(X,\beta) = \frac{1}{N}\sum_{n=0}^{N}\left[\widehat{D}(X_n,\beta) - m\,D_{dist}(X_n)\right]^2, \quad (2)$$

where $\beta$ is the learned network parameter set, $N$ is the number of pixels in the image, $\widehat{D}$ is the predicted pixelwise axon count density function, $X_n$ is the nth pixel in image $X$, and $m$ is a density scaling factor. The density scaling factor was used to increase the magnitude of the predicted pixelwise density values, allowing better regression convergence. Its value was determined during hyperparameter optimization, resulting in a final value of $m = 1000$. Since a density scaling factor was used, the trained network overestimated the density predictions by a factor of m. Thus, all density maps predicted during network application were divided by m to accurately reflect ground truth. After density map prediction, we estimated total axon count within an image as follows:

$$Axon\ Count(X,\beta) = \frac{1}{m}\sum_{n=0}^{N}\widehat{D}(X_n,\beta). \quad (3)$$

Because dataset sub-images were randomly selected from larger full optic nerve images, their edges could contain cropping artifacts such as axons that intersected the edge. Dataset images and ground truth arrays were thus padded during training and evaluation through the edge-mirroring process recommended in [15] to prevent the propagation of influence from these edge artifacts and any resulting biases in cell count. When computing the mean squared error loss function (equation 2), we did not include mirrored pixels. Training images were resized from 187 x 187 pixels to 192 x 192 pixels and extended to 224 x 224 pixels by this edge mirroring, as this size provided the optimum balance between training speed and output accuracy. Extensive data augmentation was used during training. This including image mirroring and rotation at intervals of 90° as well as random multiplicative pixel value scaling.

## D. Model Evaluation

### 1) Correction Equations

Each of the three automated counting tools cannot precisely replicate ground truth, but empirical observation shows that each tool demonstrated a relatively consistent bias, which could be corrected for. We therefore first used the validation subsets to perform the following linear bias correction, following the method established in Reynaud et al. [12]. In brief, manual counts (MC) and automated counts (AC) of axons in the validation subset were plotted against one another and fit using a linear least squares regression for each tool,

$$AC = a\,MC + b, \quad (4)$$

where coefficients a and b reflect any systematic linear bias in the estimation of MC by AC for the automated counting tool being considered. We then account for this linear bias by defining a corrected automated count, $AC_{corrected}$, as

$$AC_{corrected} = \frac{AC - b}{a}. \quad (5)$$

### 2) Statistical Analysis of Tool Performance on the Rat Image Dataset

To evaluate the three automated counting tools (AxoNet, AxonMaster and AxonJ) on the rat image dataset, we applied all three tools to the validation subsets, created correction equations as described above (equation (5)), and applied the relevant correction equation to the automated counting results from the testing subset. Differences in sub-image manual counts and the automated counts produced by each automated axon counting tool were quantified for both datasets through linear regressions, Kuskal-Wallis tests comparing the mean absolute error for each tool, and a comparison of the limits of agreement as defined by the Bland-Altman methodology [28].

In more detail, after linear regression between manual and automated counts, we examined the residual distributions from the regressions, and discovered they were not normally distributed (Shapiro-Wilk test, all p < 0.05). However, inspection of the data by histogram and Q-Q plot showed approximate normality with the exception of a small number of outliers and a slight heteroscedasticity for each distribution. In addition, linear regression is known to be robust to such slight deviations from normality, particularly in larger data sets like ours [29, 30]. We therefore judged these deviations from normality to be minor, and continued to use simple linear regression to compare model performance, taking a larger $R^2$ value to indicate a more consistent agreement between manual and automated counts.

We also calculated the mean absolute error between each automated counting tool's axon count and the gold-standard manual axon counts to quantify the accuracy for that tool. None of the mean absolute error distributions for each tool's corrected ACs were compared to a 95% confidence interval constructed from the four MCs. We defined a success rate as the proportion of images for which $AC_{corrected}$ fell within this 95% confidence interval. This approach evaluated both automated counting accuracy and precision in the same measurement.

*3) Statistical Analysis of Tool Performance on the NHP Image Dataset*

We also evaluated our rat-trained AxoNet algorithm and the two existing axon counting tools on the NHP dataset. To do so, we applied all three tools to the validation subset, created correction equations as stated above, and then applied the correction equations to the automated counting results from the testing subset. Relationships between semi-automated manual (SAM) and corrected automated counts were assessed in the same manner as they were in the rat image dataset. Since only mean axon counts were available, we were unable to compute the proportion of the automated counts that fell within the 95% confidence interval for the SAM counts or define a desired range for the limits of agreement as we did for the rat optic nerve image dataset. However, we are able to compare the limits of agreement between the corrected ACs and the SAM counts.

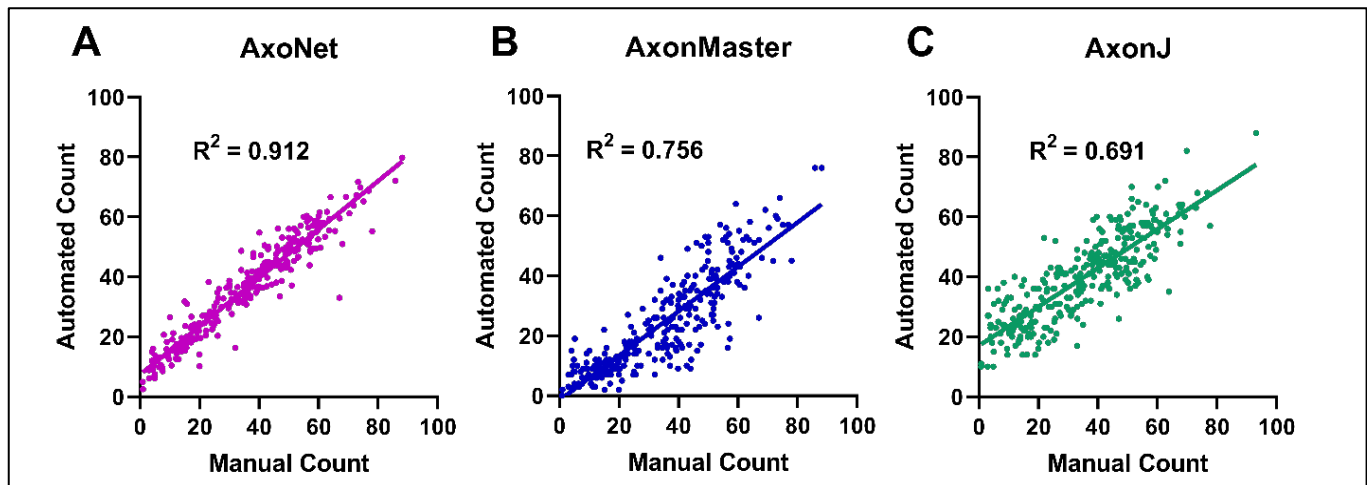

Fig. 4. Comparison between automated and manual axon counts for the rat validation subset. Comparisons are shown for AxoNet (A), AxonMaster (B) and AxonJ (C). Each data point is obtained from a single sub-image from the rat testing subset. The regression relationships between MC and AC counts were: AxoNet: AC = 0.808*(MC) + 7.3; AxonMaster: AC = 0.742*(MC) – 1.5; and AxonJ AC = 0.648*(MC) + 17.0. These relationships were used as correction equations in Figure 5.

results were normally distributed (Shapiro-Wilk: all p < 0.001), so we compared the tools' mean absolute errors using the Kruskal-Wallis test with Dunn's post hoc test.

Finally, we used Bland-Altman plots [28] to compare the limits of agreement calculated for each method. Ideally, the errors from the automated tools would lie within the range of inter-observer variability. Thus, we aimed for the limits of agreement of these Bland-Altman plots (mean count error ± 1.96•SD of count error) to be within the limits of agreement calculated for individual counters' MC relative to the mean MC. Using this definition, we computed the limits of agreement for our rat dataset as ± 14.3 axons. Additionally, for each image with four manual counters (1184 of 1474 images), the

### III. RESULTS

#### A. Rat Model Dataset Results

We first applied the three automated counting tools to the validation subset of the rat dataset to determine correction equations that accounted for linear bias, as described above (Figure 4). We then applied the automated tools to the testing subset. Before compensating for linear bias using the correction equations, the relationship between AxoNet automated and manual counts (AC and MC) in the testing subset was AC = 0.823*(MC) + 6.5 (R2 = 0.900), indicating a comparable bias to that seen when our model was applied to the validation subset. For all three automated tools, the corrected linear fit between MC and $AC_{corrected}$ resulted in regression slopes and intercepts that were not significantly different from 1 and 0, respectively (t-test for slope, $p = 0.32$, $p = 0.47$, $p = 0.41$; t-test for intercept, $p = 0.20$, $p = 0.83$, $p = 0.67$; all p-values presented in the order: AxoNet, AxonMaster, and AxonJ; Figure 5). These findings indicate that the correction equation method adequately corrected for consistent linear biases.

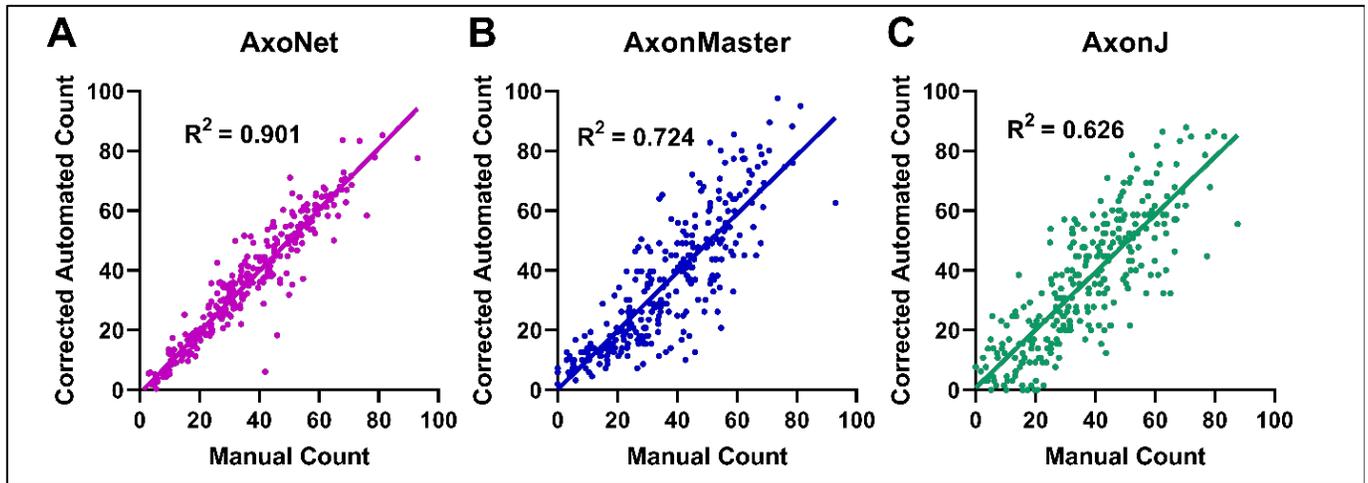

Fig. 5. Comparison between corrected automated and manual axon counts for the rat testing subset. Comparisons are shown for AxoNet (A), AxonMaster (B) and AxonJ (C). Each data point is obtained from a single sub-image from the rat testing subset. Mean absolute value predicted count errors are 4.4, 8.6, and 10.4 axons for AxoNet, AxonMaster, and AxonJ respectively. AC values are shown after applying the correction equations from Figure 4.

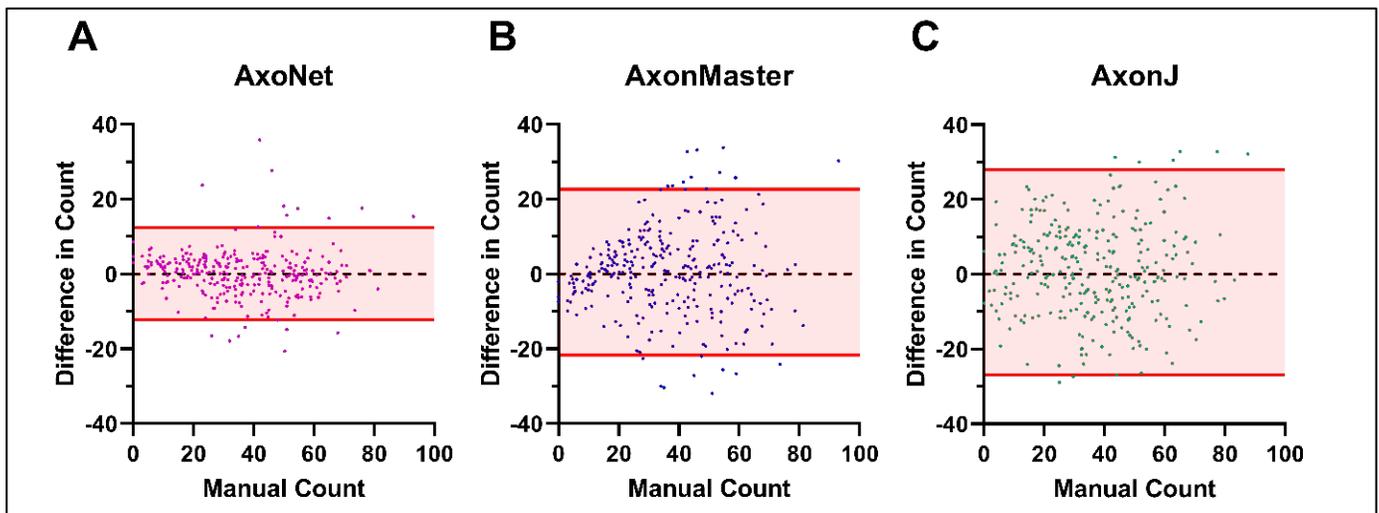

Fig. 6. Comparison of error distribution for the rat testing subset. Differences between rat testing subset MC and corrected AC are plotted against manual counts for AxoNet (A), AxonMaster (B) and AxonJ (C) as Bland-Altman plots. Each data point is a single sub-image from the rat testing dataset. Red lines represent the upper and lower bounds for the limits of agreement, calculated as mean error ± 1.96*(standard deviation of error). Limits of agreement are [-12.2, 12.8], [-21.6, 22.8], and [-26.8, 27.9] axons for AxoNet, AxonMaster, and AxonJ, respectively.

Of the three tools, AxoNet achieved the highest correlation between its corrected AC and the MC ($R^2 = 0.901$) as well as the smallest mean absolute error (Kruskal-Wallis: Chi-square = 62.58 and $p < 0.001$; Dunn's post-hoc: all $p < 0.001$, Figure 5). Only AxoNet demonstrated limits of agreement within the threshold determined by the manual count agreement (Figure 6). For the images annotated by four counters, the percentage of corrected ACs that fell within the 95% confidence interval of the manual counts was 84%, 46%, and 44% for AxoNet, AxonMaster, and AxonJ respectively. Taken together, we observe that AxoNet performed the best (i.e. the closest to manual annotations) on the testing subset of the rat dataset.

We also visualized the output of AxoNet by determining whether AxoNet was accurately replicating the density maps used during its training by comparing its predicted spatial axon count densities to ground truth (Figure 7). Generally, the density maps produced by AxoNet matched those produced by the manual annotators.

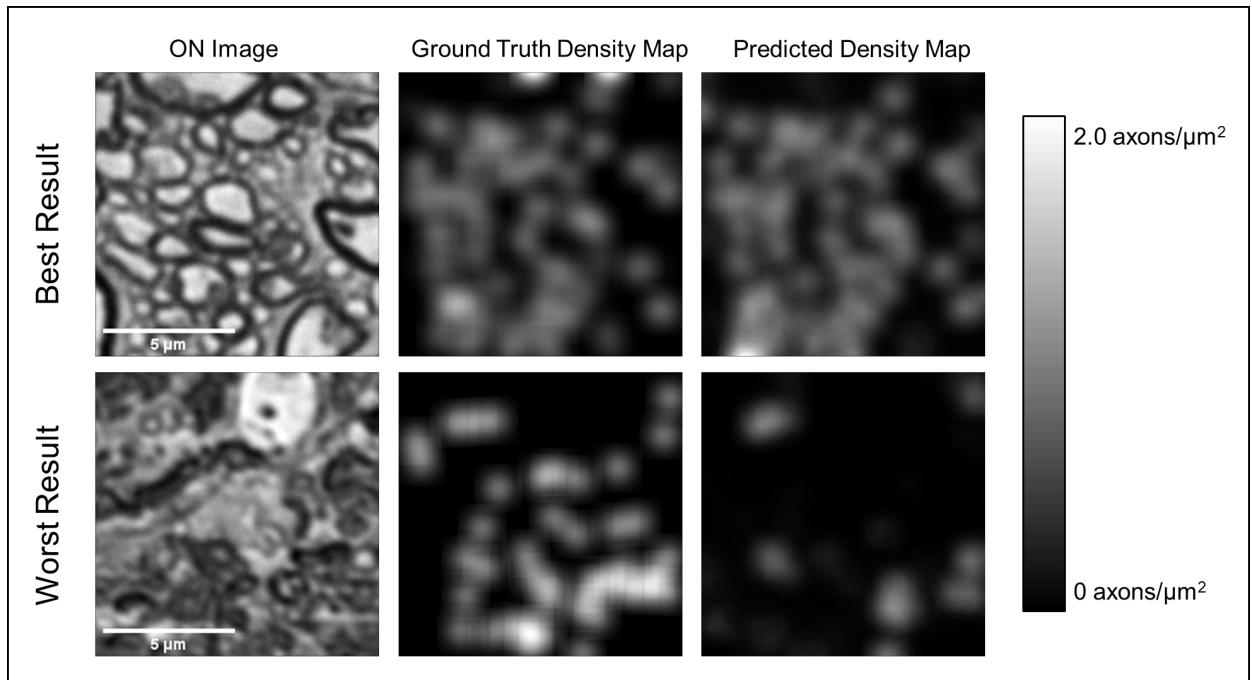

Fig. 7. Visualization of AxoNet Performance. The images from the rat testing subset which produced the smallest (top) and greatest (bottom) difference between AxoNet predicted and ground truth manual axon count are shown in the left column. The corresponding manually annotated ground truth axon count density maps are shown in the middle column, and the automatically detected axon count density maps are shown in the right column. The scale bar on the right shows the map used to visualize axon count density as greyscale intensity.

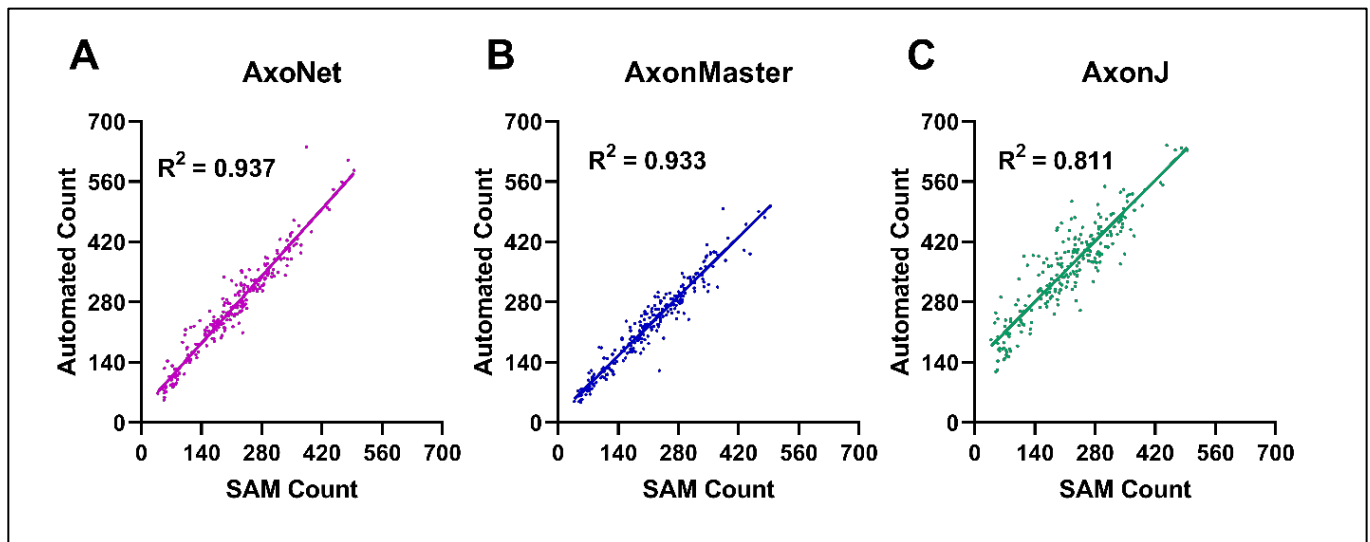

Fig. 8. Comparison between automated and manual axon counts for the NHP validation subset. Comparisons are shown for AxoNet (A), AxonMaster (B) and AxonJ (C). Each data point is obtained from a single sub-image from the NHP testing subset. The regression relationships between SAM and AC counts were: AxoNet: AC = 1.11*(SAM) + 28.5; AxonMaster: AC = 0.9849*(SAM) + 17.4; and AxonJ AC = 1.01*(SAM) + 139.2. These relationships were used as correction equations in Figure 9.

*B. NHP Model Dataset Results*

We then applied these three automated counting tools to the NHP dataset. We first assessed the performance of the three tools using the validation subset of the NHP dataset in order to construct bias correction equations relating each tool's AC to the SAM count. When applied to the validation subset of the NHP dataset, AxoNet achieved a higher correlation between SAM count and AC than the other two tools, although AxonMaster needed less bias correction (Figure 8), likely because it had been optimized for the NHP dataset.

The automated counting methods and their correction equations were then applied to the testing subset of the NHP dataset to directly compare their ability to accurately quantify the number of axons present in each image. For all three automated tools, the corrected linear fit between SAM count and $AC_{corrected}$ resulted in regression slopes and intercepts that were not significantly different from 1 and 0, respectively (t-test for slope, p = 0.98, p = 0.47, p = 0.81; t-test for intercept, p = 0.67, p = 0.82, p = 0.71; all p-values presented in order: AxoNet, AxonMaster, and AxonJ; Figure 9). Of the three tools, AxoNet achieved the highest correlation between its corrected automated and manual counts ($R2=0.945$), with AxonMaster achieving a comparable correlation ($R2=0.938$). AxoNet and AxonMaster both had lower mean absolute error when compared to AxonJ (Kruskal-Wallis: Chi-square = 154.1 and p < 0.001; Dunn's post-hoc: both p < 0.001, Figure 9), while AxoNet and AxonMaster had similar mean absolute error values to one another (p = 0.976). AxoNet and AxonMaster produced comparable limits of agreement, whereas AxonJ's limits of agreement were larger (Figure 10). We packaged AxoNet into a user-friendly plugin for Fiji and ImageJ. This plugin is capable of counting full rat optic nerve images in about 15 minutes (Figure 11). We typically count c. 80,000 "normal"-appearing axons in a healthy nerve, consistent with previous reports [5, 6, 31].

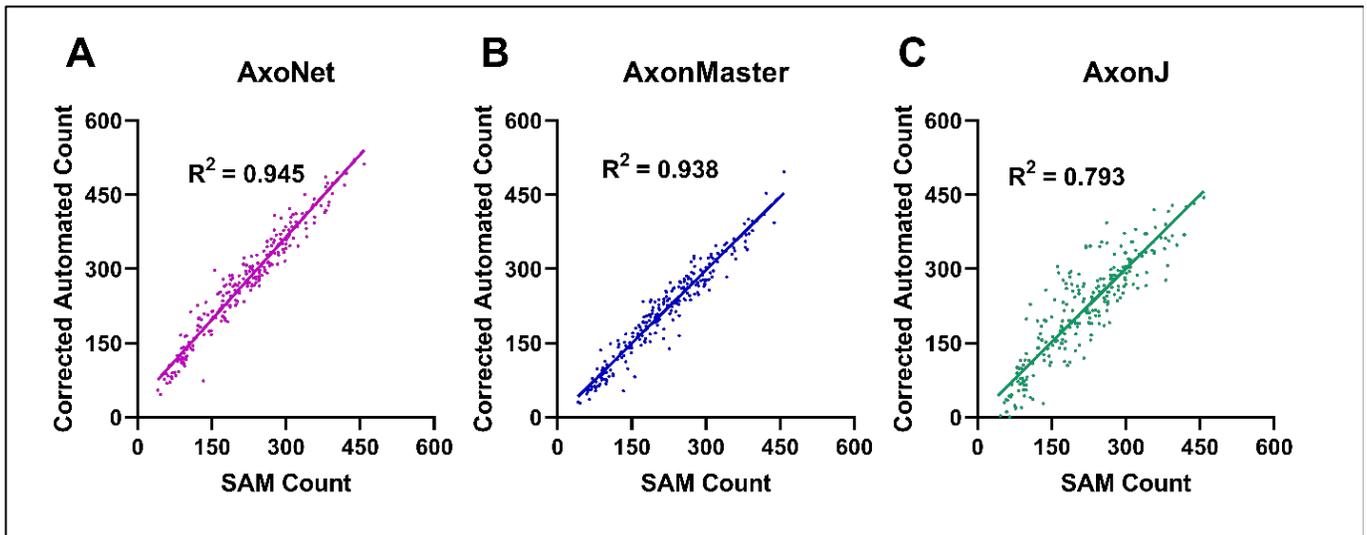

Fig. 9. Comparison between corrected automated and manual axon counts for the NHP testing subset. Comparisons are shown for AxoNet (A), AxonMaster (B) and AxonJ (C). Each data point is a single sub-image from the NHP testing subset. Mean absolute predicted count errors are 17.8, 18.2, and 35.0 axons for AxoNet, AxonMaster, and AxonJ respectively. Automated count values are shown after applying correction equations in Figure 8.

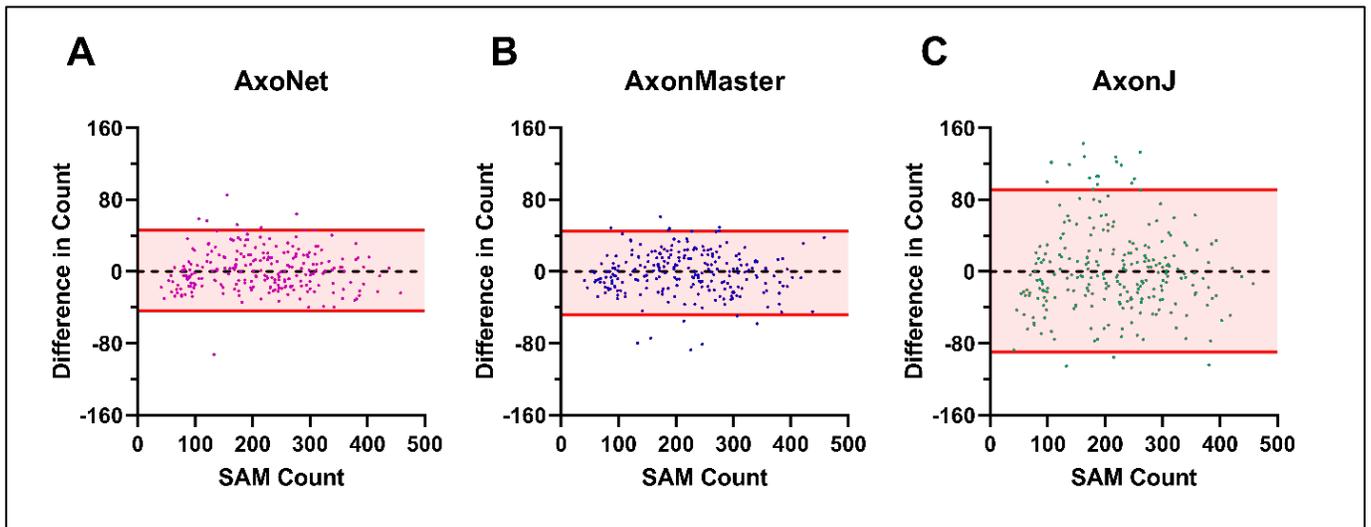

Fig. 10. Comparison of error distribution for the NHP testing subset. Differences between NHP testing subset semi-automated manual count and corrected AC are plotted against semi-automated manual count for AxoNet (A), AxonMaster (B) and AxonJ (C) as Bland-Altman plots. Each data point is a single sub-image from the rat testing subset. Red lines represent the upper and lower bounds for the limits of agreement, calculated as mean error ± 1.96*(standard deviation of error). Limits of agreement are [-43.3, 45.7], [-48.4, 44.6], and [-89.8, 91.0] axons for AxoNet, AxonMaster, and AxonJ respectively.

IV. DISCUSSION

The purpose of this study was to develop and evaluate a new approach to automatically count "normal"-appearing RGC axons in a diverse dataset of healthy and damaged optic nerve cross sections. Such an automated axon counting tool is a useful tool in studying glaucoma and potentially other neurodegenerative disorders. We designed this new approach to work well over a range of image qualities and for multiple mammalian species. AxoNet's predicted axon counts proved to be highly correlated to manual axon counts in both the rat and NHP datasets, indicating that it met our requirements for an automated axon counting tool. As judged by the uniform error over the range of manual axon counts (Figures 6 and 10), AxoNet performed equally well on images of damaged vs.

healthy optic nerves. This is significant because axon counting is more difficult in diseased tissue, and suggests promise for the use of AxoNet as a tool for nerve damage analysis in experimental glaucoma.

Prior to building AxoNet, we explored the methodologies previously used to create existing automated axon counting tools. AxonMaster uses a fuzzy c-means classifier as an adaptive thresholding method to segment axon interiors from the darker myelin sheath. These clusters are then filtered by size and circularity before counting axons. AxonJ uses a Hessian operator to identify the darker myelin sheath and then performs similar adaptive thresholding and connected region size filtering region before counting the connected regions as axons. When applied to the rat dataset, these two tools produced adequate segmentation of total axon area in optic nerve images, but often did not produce accurate segmentation of individual axons, leading to inaccurate counts. We also attempted to apply two other segmentation techniques, ilastik [32] and the basic pixel segmentation U-Net [15]. These synthesis between a convolutional neural network architecture designed for cell segmentation, the U-Net, and a count density prediction strategy. This method avoids the hard problem of axon segmentation in lower-resolution light microscopy, trading the ability to analyze single-axon morphology for the most accurate axon count.

This study was limited by several factors. First, and most important, to date AxoNet has been trained to count only "normal"-appearing axons, similar to existing axon-counting software. The classification of an axon as "abnormal" in appearance does not necessarily imply that the axon is non-functional, and thus our tool may not count axons that are in fact conducting visual information. However, due to AxoNet's generalizability and lack of reliance on hand-crafted features specific to "normal"-appearing axons, it can be extended to count or even segment other features of both healthy and glaucomatous optic nerves, such as glial processes, nuclei, "abnormal" axons, large vacuoles, and extracellular matrix. We are currently extending AxoNet to quantify these features.

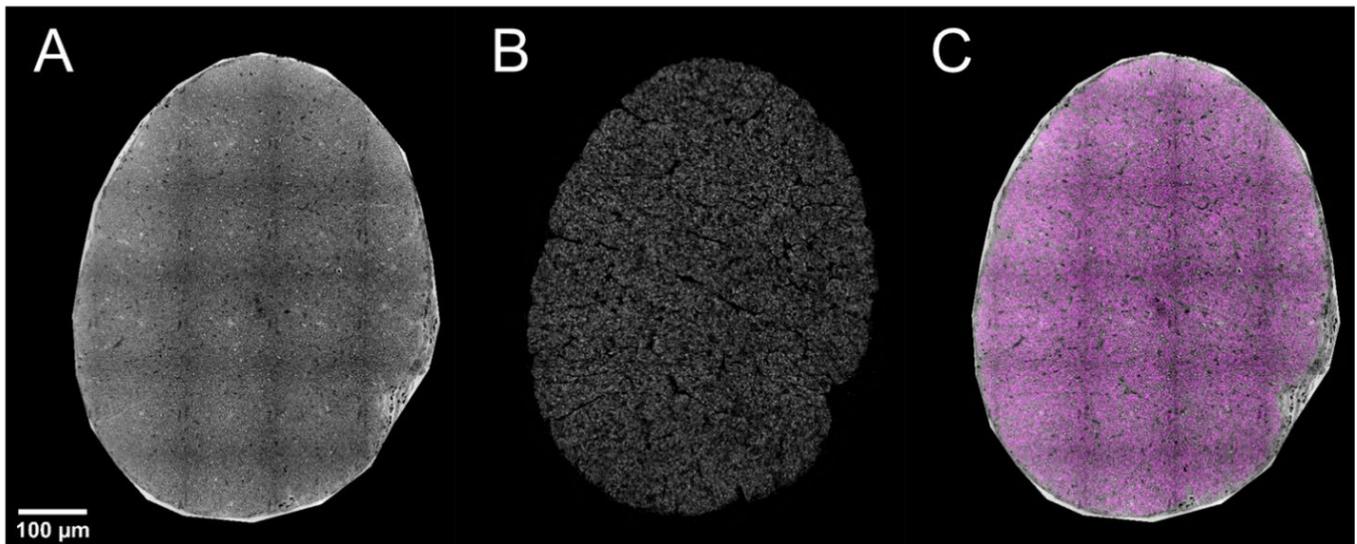

Fig. 11. AxoNet Plugin Results. After using the AxoNet plugin for ImageJ and Fiji on an image of a full rat optic nerve (A), the output axon density map (B) and the combination of these two images (C) are displayed. The combination of these two images is shown with the input image (A) in greyscale and the axon density map (B) overlaid in pink. Axon density scale is not provided here because these full images are scaled down significantly for inclusion in the manuscript and color scale is indistinguishable at this resolution. A grid of dark lines is visible in panel A; these lines correspond to tile edges from the microscopy imaging and are an artifact of visualization only since counts are carried out on much smaller portions of the full image.

approaches also resulted in inaccurate counts, especially when applied to damaged tissue; therefore, we adapted an alternate cell counting framework introduced by Lempitsky et al. [16]. This approach avoids the difficult task of semantic segmentation and instead predicts a pixelwise cell count density estimate. The authors accomplished this through using machine learning with hand-crafted pixelwise features [16]. More recently, attempts have been made to perform similar count density function estimations using convolutional neural networks [33] and adapted U-Net architectures [34] for crowd counting, which is a technically similar problem to cell counting. Convolutional neural networks have also been used recently for axon segmentation in scanning and transmission electron microscopy images of mammal and human spinal cord [35]. The tool produced in this work is the result of this

Such analysis of features beyond "normal"-appearing axons may provide new insight into the pathophysiological processes of glaucomatous nerve degeneration.

Second, the linear bias correction equations determined in this study were suitable for our data set, but may not necessarily be accurate for other data sets, since the conditions which create these systematic biases may vary with experimental treatment, imaging protocols, or tissue processing protocols. However, we do not expect such effects to be severe, since we intentionally included these sources of variability within the two image datasets used in this study and AxoNet still performed well. Nonetheless, it would be prudent to calibrate AxoNet for each new application, which can be done through using correction equations like those created with our validation subsets or network retraining with a new dataset according to the training protocol detailed above.

A third limitation is that all manual counts were conducted by members of one lab, and it is possible that manual counts generated in different research groups could be slightly different from ours since manual counting itself is not entirely unambiguous. This uncertainty is inherent in axon quantification and cannot be avoided, although to enhance repeatability we have explicitly described our definition of "normal"-appearing axons and have made the training data publicly available.

Presently, AxoNet regresses a pixelwise count density function which is integrated over the full image to return a count. Fitting the density function is accomplished through the minimization of a mean squared error loss function evaluated at each pixel (Equation 2). This loss function may be overly sensitive to zero-mean noise and other variations in training images. Lempitsky et al. [16] originally solved this problem through the Mesa loss function, which used a maximum subarray algorithm to find the image region with the largest difference between automated and manual counts and minimized count error over this region instead of at every pixel [16]. When we attempted to use this loss function during our training, the resulting method was far too computationally expensive and resulted in a prohibitively long training time (on the order of hours per training step). However, developing a new loss function which avoids computing the mean square error at every pixel per iteration but does so without the computational expense may improve AxoNet's performance in terms of accurate axon count insensitive to image noise.

The successful use of the rat-trained AxoNet to count NHP images is indicative of the versatility of our method, even without re-training. However, the network can be easily re-trained on a new counting case if needed. If there is adequate training data in the new set, the deep learning framework can adapt itself to new applications without requiring any changes in handcrafted features. Data augmentations like those described in the methods can be applied to improve network learning from limited datasets, as was done in the first published application of the U-net architecture [15].

We can also use AxoNet to count axons in full rat optic nerve images by subdividing the full image into tiles for individual processing. This tile-based processing was necessary because of the prohibitive computational expense involved in applying the U-net architecture to large images. However, tile-based processing has the potential to create edge artifacts by cutting off portions of cells on the borders of each tile. We correct for this potential error by padding the edges of each processing tile with bordering pixels from adjacent processing tiles. Including this information from bordering tiles meant that the resulting density map prediction was not affected by these potential tile cropping artifacts. Once processed, the resulting density map was cropped back to its original tile size. This padding was not done when it would have required pixels from beyond the image boundaries. These padded tiles were then also mirrored, as described for model training above.

When running on the system used for this study (Windows Desktop, Intel i7-3770 CPU at 3.40 GHz, 16 GB RAM; Dell, Round Rock, TX), AxoNet counts the axons within a full rat optic nerve image in approximately 15 minutes. For comparison, it took AxonJ and AxonMaster approximately 30 minutes and 1 hour to count the axons within a full rat optic nerve image. Therefore, our tool can be applied to analyze full optic nerve images with runtimes comparable to, or better than, those of the existing automated tools.

## V. CONCLUSION

We have successfully applied a deep learning method to accurately count "normal" axons in both rat and non-human primates, and in both healthy and experimentally glaucomatous optic nerve sections. Additionally, we have compared AxoNet to two previously published automated counting tools and shown that AxoNet performs as well as or better than these two tools in counting healthy axons in these two datasets. Our tool is available online as an ImageJ plugin and can be installed by following the instructions at https://github.com/ethier-lab/AxoNet-fiji. The code and data we used to train the model can be found at https://github.com/ethier-lab/AxoNet.